\documentclass{aa}
\usepackage{graphicx}
\usepackage{txfonts}
\usepackage{epsfig}
\usepackage{times}
\def\ltsima{$\; \buildrel < \over \sim \;$}
\def\gtsima{$\; \buildrel > \over \sim \;$}
\def\simlt{\lower.5ex\hbox{\ltsima}}
\def\simgt{\lower.5ex\hbox{\gtsima}}

\begin{document}
   \title{A model for the X-ray absorption in Compton--thin AGN}
   \author{Alessandra Lamastra, G. Cesare Perola and Giorgio Matt}
   \offprints{}
   \institute{Dipartimento di Fisica ``E. Amaldi'', Universit\`a degli Studi Roma Tre,
via della Vasca Navale 84, I-00146 Roma, Italy}
   \date{Received ; Accepted }
   \abstract{ The fraction
of AGN with photoelectric absorption in the X-rays ranging from N$_H$ of 10$^{22}$ up to about 10$^{24}$  cm$^{-2}$ (Compton-thin) appears
observationally to be anticorrelated to their luminosity $L_x$. This recently
found evidence is used to investigate the location of the absorbing
gas. The molecular torus invoked in the unified picture of AGN, while
it can be regarded as confirmed on several grounds to explain the
Compton-thick objects, do not conform to this new constraint, at least
in its physical models as developed so far. In the frame
of observationally based evidence that in Compton-thin sources the
absorbing gas might be located far away from the X-ray source, it is
shown that the gravitational effects of the black hole (BH) on the molecular gas in a
disk, within 25-450 pc (depending on the BH mass, from 10$^6$ to
10$^9$ M$_{\odot}$),
leads naturally to the observed anticorrelation, under the assumption
of a statistical correlation between the BH mass and $L_x$. Its normalization
is also reproduced provided that
the surface density, $\Sigma$, of this gas is larger than about 150-200 $M_{\odot}$ pc$^{-2}$, 
and assuming that the bolometric luminosity is 
one tenth of the Eddington limit. Interestingly, the required
values are consistent with the value of the 300 pc molecular disk in our own galaxy,
namely 500 $M_{\odot}$ pc$^{-2}$. In a sample of nearby
galaxies from the BIMA SONG survey, it is found that half of the objects have central $\Sigma$
larger than 150 $M_{\odot}$ pc$^{-2}$. Given the simplicity of the proposed model, this 
finding is very encouraging, waiting for future higher resolution surveys in CO
on more distant galaxies.
   }
\authorrunning{A. Lamastra, G.C. Perola and G. Matt}
\titlerunning{Compton--thin type 2 AGN}
   \maketitle


\section{Introduction}
Similar to dust extinction in the optical, photoelectric absorption in 
X-rays provides clues on the environment of Active Galactic Nuclei (AGN). 
In general the latter provides information which is more straightforward 
to interpret than the former, because it depends a) on atomic properties and 
consequently only on the chemical composition of the medium, b) on the adoption 
of a shape for the X-ray continuum, which is empirically known to follow a 
simple Power Law (PL). The values of the 
spectral photon index, $\Gamma$, display a rather modest dispersion (e.g. Perola et al. 2002; 
Piconcelli et al. 2005); 
furthermore, in any single object under consideration, and provided that the absorption is
Compton-thin (i.e. with N$_H$ not exceeding $\sigma_{Th}^{-1}$=1.5$\times$10$^{-24}$
cm$^{-2}$), the exact value of $\Gamma$ 
can be directly measured, given an adequately wide observational band and sufficient 
statistics for the spectral counts. Here, though, it is more important to stress 
the limitations, which are 
mainly of three sorts:\begin{enumerate}
\item The source of the PL photons is, for most
practical purposes,  point-like,
and therefore it remains questionable how to relate the total
absorbing matter density (the absorbing column N$_{H}$) to bi-dimensional information
(either direct or indirect) on the circumnuclear matter from observations in 
other wave-bands. In other words, usually it cannot be excluded
that the observed column might be pertaining 
to that unique line of sight, rather than to the overall configuration of
the circumnuclear matter.
\item While it is spectroscopically possible (as amply demonstrated even before
the use of very high resolution instruments) to disentangle the imprints of a
multi-phase (in ionization terms) absorber, it is by no means immediately 
possible to recover the space distribution of the phases along the line of sight. 
When the ionization is attributed to the photoelectric effects of the
PL source as a reasonable hypothesis (better still when it can be observationally 
proven not to be of collisional origin), it is generally regarded as most likely 
that the high ionization gas is very close to the centre. This leaves however 
quite open the possibility that the low (in practice, indistinguishable from neutral) 
ionization gas might lie either much further away or equally close but dense 
enough to remain neutral.
\item When N$_{H}$ is larger than about $10^{24}$ cm$^{-2}$, the column is Compton-thick, that is a diffusive process
adds to the absorption process to deplete the specific intensity along the
line of sight. A truly thick absorber ($\tau_{Th}\gg$ 1) reduces the intensity by
several orders of magnitude all the way throughout the hard X-ray band, and renders
in fact its quantitative evaluation impossible, except for a ``lower limit" (e.g. Matt et 
al. 1999).
\end {enumerate}

It must be emphasized that, in general, AGN
spectra contain another component in addition to the PL, which becomes evident
beyond 8-10 keV. It is due to reflection from Compton-thick, neutral gas,
and, as expected on physical grounds, it is accompanied by a strong fluorescent
iron line. From a pure observational point of view, this component is often the only visible one below $\sim$ 10 keV when the PL is photoelectrically absorbed by N$_{H}$ of about $10^{24}$ cm$^{-2}$
(corresponding to a turnover in energy at about 10 keV), or greater. Its intensity is typically a few percent of the primary one in the 2-10 keV band. In such systems,
it is regarded as very reasonable to assume that the absorbing and 
the reflecting gas belong to one and the same circumnuclear structure. Historically, 
it was the spectropolarimetric measurements in the optical band of NGC 1068
which led Antonucci \& Miller (1985) to propose the so-called ``unified model". This model
(Antonucci 1993) envisages the existence of a thick 
and dusty ``torus" of molecular gas around the nucleus, extending out to a few tens of pc
at most (Risaliti et al. 1999)
and with a substantial covering factor to account for the large 
ratio of type 2 to type 1 Seyferts, at least in the local Universe 
(4:1, if also Seyfert 1.8 and 1.9 are included: Maiolino \& Rieke 1995; Ho et al. 1997; it
is worth recalling that
the local Universe is dominated by relatively low luminosity objects as compared to the 
Quasi Stellar Objects, QSO). In its basic form, the model has been widely confirmed, most notably
by measurements in the hard X-ray band, which showed that most type 2 Seyferts are X-ray
absorbed. The situation, however, is likely to be more complex than the zero-th order
model would predict. On one hand, there is increasing evidence of 
a substantial population of optically inactive, obscured AGN, both in the local
(Maiolino et al. 2003) and more distant (Comastri et al. 2002; Brandt \& Hasinger 2005) Universe.
This implies that the ratio bewteen X-ray obscured to unobscured AGN may be larger than 
that between type 2 and type 1 Seyferts.
On the other hand, other absorbing regions,  besides the torus, may exist in the
complex environment of AGN: dust lanes (Malkan et al. 1998), 
starburst regions (Weaver 2002), the galactic disk (Maiolino \& Rieke 1995).
Indeed, it has been suggested (see e.g. Matt 2000, 2004 and references therein)  
that Compton--thick absorption (which, in the local Universe, 
is observed in about half of optically selected type 2
Seyferts, Risaliti et al. 1999, Guainazzi et al. 2005, the percentage rising when ``elusive''
AGN are also taken into account, Maiolino et al. 2003) is related to the torus, while 
Compton--thin absorption is due to one (or more) of the other possible absorbing regions. 
(If a dust/gas ratio typical of the cold interstellar medium of our Galaxy is adopted, 
Compton--thin absorption, i.e. N$_{H}$ in the 10$^{22}$--10$^{24}$ cm$^{-2}$ range,
is sufficient to explain the extinction of the BLR lines). 

This paper will address the question of the location of the absorber
in the light of new results from X-ray surveys in the 2-10 keV band. The sensitivity reached 
in this band with the XMM-Newton and Chandra satellites made it, for the first time, 
possible to study AGN of both type 1 and 2, with sufficient statistics to investigate 
the  N$_{H}$ distribution as a function of the intrinsic (PL) luminosity 
and of the cosmic epoch out to z about 4.

Surveys limited to 10 keV on the high energy side (and with instrument sensitivity 
dropping rapidly beyond 5-7 keV) are doomed to miss almost completely the 
Compton-thick objects. Thus the results must be read as pertinent, among the AGN 
classified through optical spectroscopy as type 2, only to those 
with  N$_{H}$ less than $10^{24}$ cm$^{-2}$. Basically, the question addressed in this paper 
is whether the new results can be understood in the context of the unified model 
based on just one element, the torus mentioned above, or, if not, if one can conceive 
of just one realistic (astrophysically) additional element able to provide a positive
answer, with the least possible number of assumptions. In particular, 
the new contribution of this paper is the use of one further
constraint, namely the decrease of the fraction of Compton-thin obscured AGN with increasing X-ray luminosity,
an anticorrelation which have recently emerged from the statistical
analysis of the abovementioned types of survey (Ueda et al. 2003; La Franca et al. 2005).

In Sect. 2 this new constraint is described.
In Sect. 3 the current physical models of the torus are discussed and shown to 
be unable to account, even qualitatively, for the anticorrelation.
 In Sect. 4 it is shown that, if the obscuration is due to molecular gas in galactic disks, then the gravitational effects of the black hole on its 
spatial distribution within 25--450 pc (depending on the mass) leads
naturally to an anticorrelation. An encouragingly reasonable quantitative 
match is achieved under
two conditions: that the molecular mass surface density exceeds
$\sim$150 $M_{\odot}$ pc$^{-2}$ within 25-450 pc, and that in a statistical sense
the bolometric luminosity is proportional to the black hole mass,
and equal to about 10\% of the Eddington luminosity. In Sect. 5
these two conditions are discussed in the light of the evidence
available, and a positive conclusion is drawn.

\section{A new observational constraint: the luminosity--absorption anticorrelation}
A statistical investigation on the photoelectric absorption properties
of Compton-thin AGN  was first obtained, down to a flux limit
F(2-10 keV) of about 10$^{-13}$ erg cm$^{-2}$ s$^{-1}$ (which corresponds to about 25$\%$ 
of the X Ray Background, XRB) with imaging instruments operating up to 10 keV 
onboard ASCA and BeppoSAX (e.g. Ueda et al. 1999; Fiore et al. 1999). 
The main problem with these pilot surveys
is the error boxes at fluxes close to the limit, whose size left
substantial uncertainties on the optical identification, especially for the 
type 2 AGN (see La Franca et al., 2002, for a discussion on this point).
The situation improved greatly in the last four years thanks to Chandra
and XMM-Newton, both of which image the sky up to 10 keV with much higher
sensitivity and much better angular resolution (at the arcsecond level in the
case of Chandra). 
 The combination of these two qualities have led to positionally based identifications,
followed by spectroscopic identifications of the counterparts, which are 
regarded as fairly secure down to magnitudes R about 24.
The X-ray flux limit achieved with a high fraction of identified sources
is approximately 10$^{-15}$ erg cm$^{-2}$ s$^{-1}$ (e.g. La Franca et al. 2005).
With the instrument sensitivity covering the spectra down
to 0.5 keV, along with the knowledge of $z$, it is possible
to estimate both the intrinsic value of  N$_{H}$ and the X-ray luminosity L$_{x}$ 
corrected for absorption. After accumulating a sufficiently large
sample to cover the L$_{x}$-$z$ plane, one can investigate simultaneously
the X-ray luminosity function of AGN (irrespective of their being optically
classified as type 1 or 2), its cosmic evolution, and the distribution
of N$_{H}$ as a function of L$_{x}$ and $z$. In the following, L$_{x}$ will be 
referred to as the intrinsic luminosity in the 2-10 keV band.
As noted in Sect. 1, truly Compton-thick sources remain basically
unsampled, thus the constraint emerging from these studies apply
only to sources with N$_{H}$ ranging from (nominally) zero to about
$10^{24}$ cm$^{-2}$.
With the statistics achieved so far it appears premature to
observationally quantify the functional form of the  N$_{H}$ distribution.
At present it is convenient to adopt a conservative approach,
by using the ratio $\xi$ = (number of sources with log (N$_{H}$) $>$ 22)/(total 
number of sources).
Ueda et al. (2004) have shown that $\xi$ is dependent on L$_{x}$,
that is $\xi$ = $\xi$(L$_{x}$), where $\xi$ decreases with increasing L$_{x}$.
Moreover, the same authors reached the conclusion that
this property appears to be quantitatively independent from z.
Recently, with a larger sample, which overlaps the one utilized by Ueda
et al. (2004), but yields a wider coverage of the L$_{x}$-$z$ plane,
La Franca et al. (2005) confirmed the first finding, but not the second.
Namely they find evidence of a $z$ dependence, thus formally $\xi$=
$\xi$(L$_{x}$, $z$), where $\xi$ increases and becomes shallower as $z$ increases.

In this paper only the dependence of $\xi$ on L$_{x}$ in the local universe,
as obtained by La Franca et al. (2005) from $z$=0 to $z$=0.1, 
will be used as
the novel constraint, while the dependence on $z$ will be briefly commented upon
and left as the subject of further investigations.

The next Section is devoted to investigate whether this anticorrelation
can be understood in the frame of physical models of the torus.

\section{The case of the ``thick torus"}
Several authors (Krolik $\&$ Begelman 1986, 1988, 
Beckert $\&$ Duschl 2004, Vollmer, Beckert $\&$ Duschl 2004, Beckert et al. 2004) 
have addressed in physical terms the dynamical
and geometrical structure of the "thick torus", which 
is assumed to be cylindrically symmetric. If $R$ is the distance from the
AGN centre (the site of the black hole, BH, with mass $M_{BH}$), and $H(R)$ is the
scale height (vertical geometrical thickness), these models are often
depicted (see Fig. 1) as if the "torus" were sharply confined within
2$H(R)$, with all the obscuring matter within and nothing outside. This
type of descriptive picture applies well to the separation of the
Compton-thick sources from the rest. One could imagine a smoother
transition of the obscuring matter density, in order to accomodate
also the absorbed, but Compton-thin sources. The question is then the
following: do these models, on the grounds of their physical
principles and self-consistency, behave in a way that, without
additional ad hoc assumptions, comply with the constraints 
described in Sect. 2?
\begin{figure}[h]
\begin{center}
\includegraphics[width=7 cm]{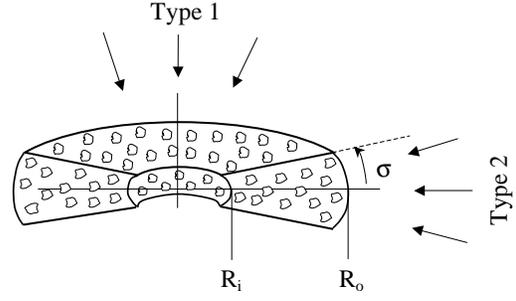}
\caption{Geometry of the "thick torus" (Nenkova et al. 2002) }\label{}
\end{center} 
\end{figure}
In these models $H$ is an increasing function of $R$, simply as a consequence
of the gravity exerted by the BH, in addition to that exerted by a concentric
and, in first approximation, spherically symmetric stellar structure.
In order to ensure the required scaleheight together with the most
effective obscuration, the matter is assumed to be in the form of
dense molecular clouds, endowed with the adequately large velocity
dispersion. The clouds are considered to be either self-gravitating
(Beckert $\&$ Duschl 2004, Vollmer, Beckert $\&$ Duschl 2004) or in pressure equilibrium with a hot phase. 
The clouds can closely interact with each other or be subject to tidal shear,
both effects leading to their destruction: they set stringent limits
on the cloud size (hence their internal density) as well as on their
number density as a function of $R$. Beckert et al. 
(2004) have shown that the geometrical thickness of the torus is given by:

\begin{equation}
H = \sqrt{{2R^3\dot{M} \over c_s M(R) }}
\label{thicknesstorus}
\end{equation}

\noindent
where $\dot{M}$ is the total mass accretion rate, $M(R)$ the mass within the radius
$R$ and $c_{s}$ is the sound speed. 

In order to maintain the obscuration effective as required by observations 
(see Sect. 1), $\dot{M}$ must be  much larger
than $\dot{M}_{Edd}$, namely most of the matter contributing to $\dot{M}$
must be lost, and only a fraction of it ($\dot{M}_{lum}$) 
is allowed to feed the AGN power (see Fig. 4 in Beckert $\&$ Duschl 2004.).
If $\dot{M}_{lum}$ increases with $\dot{M}$, which seems reasonable to assume,
$H/R$, hence the covered fraction of 4$\pi$, would increase with the luminosity,
contrary to the novel constraint. 

One might of course read this conclusion as suggesting that something, 
to be physically demonstrated, is acting in the system to ensure 
anticorrelation between $\dot{M}$ and $\dot{M}_{lum}$.
So far no account has been taken of the feedback on the state of
the matter due to the irradiation by the AGN itself.
The ionization-evaporation effects set a limit on the minimum
distance to which an absorbing cloud can survive. A useful 
quantification follows (Krolik $\&$ Begelman 1988) from equating the 
``evaporation time--scale'' and the ``accretion time--scale'':
\begin{equation}
R_{evap}=\frac{0.79 M_{gas,5}L_{44}}{\dot{M}_{torus}R_{out}N_{cl,24}T_{5}^{1/2}}   pc
\end{equation}

\noindent
where L$_{44}$ is the central ionizing luminosity in units of 10$^{44}$ erg s$^{-1}$,
$M_{gas,5}$ is the gas mass in units of 10$^5 M_{\odot}$,
 $N_{cl,24}$ is the column density of a single cloud in units
of 10$^{24}$ $cm^{-2}$, $T_5$ is the temperature at the sonic point of the evaporating flow
in units of 10$^5$ K, and $R_{out}$ is the outer radius of the configuration.
In a purely geometrical model of the ``torus'', described as
a cylinder with a cavity (Lawrence 1991), the radius of the latter
is bound to increase with the luminosity. Consequently, the
covered fraction of 4$\pi$ would decrease. However this would be
an acceptable answer to the anticorrelation described in Sect.2 if one were
able to demonstrate that physically $H/R$ can be maintained constant (or decrease)
versus $R$. The physical models summarized above, if $M(R)$ increases with $R$ less
than linearly (as it is the case when this term is dominated by the BH mass)
 show instead that
the pure effect of the BH gravity involves a concave geometry (see Fig.~3 in
Beckert \& Duschl 2004),
hence the covered fraction is basically determined by $H$ at $R_{out}$.
This implies that the fraction of obscured sources would be independent
of the luminosity up to a certain value at which the illumination is so strong to forbid the
formation of the torus. This is thence an on-off effect,
which could be relevant to the issue of the fraction of Compton-thick 
sources as a function of their luminosity: this issue at present
lacks an observationally secure point of reference, and will not be further discussed
in this paper.

\section{Obscuration by molecular gas in the galactic disk}

On the basis of a statistical analysis of the correlation between
the spectroscopic optical classification of nearby AGN and the inclination
to the line of sight of the host (typically spiral) galaxies,
Maiolino \& Rieke (1995) found the following. For the extreme type 2
there seems to be no correlation. On the contrary, types 1+1.2+1.5
are mostly found in face-on galaxies, and the intermediate types, 1.8+1.9, are
mostly found in edge-on galaxies.
They therefore suggested that the obscuration in the
intermediate AGN types is associated with the gas in the galactic
disks. Risaliti et al. (1999) showed that these types, when observed in X-rays,
correspond to Compton-thin AGN. 

It remains to be seen if the anticorrelation discussed in Sect. 2.
could be explained within this scheme.
The gas content at large in the disk of spiral galaxies correlates
with their morphological type. In particular, the {\it mean} HI surface
density over the disk out to the same isophotal radius is about twice 
in Sc that in Sa galaxies (Roberts \& Haynes, 1994). 
However, the concentration of the gas is generally
associated with the spiral arms. So, it should increase towards the center
in the Sa type, where the arms are more tightly wound, 
and decrease in later types.
Furthermore, since the far infrared emission (FIR)
appears to correlate with the CO flux, thus with the molecular
gas content, it is noteworthy that the {\it mean} FIR surface density
declines slightly from the Sa to the Sc type (Roberts \& Haynes, 1994).
It should be therefore a fair approximation to assume that the relevant,
HI + H$_2$, gas content
in the innermost, say 0.5 kpc scale, region is not significantly
(for our purpose) dependent on the morphological type.
On the other hand, along the sequence
Sa-Sc, the bulge component relative to the disk is largest in the
earliest type and decreases thence. In addition, there exist
a fairly close relationship
between bulge luminosity and central BH mass (Ferrarese \& Ford 2005).
Thence, if the AGN $L_{x}$ were (statistically) correlated with $M_{BH}$,
one should not expect the existence of the trend given by $\xi (L_x)$

This statement follows, of course, only if one refers to the gas
content alone, and the gravitational influence of the BH (and
the bulge) on the geometrical thickness of the gas within the
first few hundred parsec from the center is disregarded. In the
following, it is argued that this influence might indeed shed some
light on the anticorrelation that one would like to explain.
To illustrate the point, an ``idealized'' design is adopted, where
the molecular gas is confined within a self-gravitating, starry galaxy disk,
whose surface density $\Sigma$ remains practically constant within a
distance from the center of the order of one kpc. The scale height
$H_c$ of the molecular clouds is regulated by the disk gravity and their 
velocity dispersion. When closing inwards, the bulge and the 
BH gravity adds to that of the starry disk, inducing a progressive
shrinking of $H_c$. In geometrical terms this picture is quite
similar to the one which applies to the physical models of the
torus, summarized in the previous Section. In physical
terms it is simpler, because outside a certain radial distance
the collisional and tidal effects have no serious impact and can
be disregarded.  

The picture is characterized by the following parameters.\begin{itemize}
\item The number density $n_0$ of the clouds at Z=0.
\item The scale heigth defined according to
\begin{equation}
          n(Z) = n_0 e^{-({Z \over H_{c}})}
\end{equation}
\item The clouds idealized as single spherical entities. Those which
matter most for the absorption have the largest mass, i.e. the clouds that in
our Galaxy dominate the mass function (where most of the mass resides).
Thus single values for mass, radius, internal
density are used: $m_{c}$, $r_{c}$, $\rho_{c}$. Further parameters are the velocity 
dispersion $\sigma_{c}$ and the internal temperature $T_c$.
\item The clouds are assumed to be in pressure equilibrium with an external gas
(namely  $m_{c}$ $<$  $m_{jeans}$), whose $\rho_{e}$ and $T_{e}$ are such that
$\rho_{e}T_{e}$ = $\rho_{c}T_{c}$, and $H_{e}$ $\propto$ $H_{c}T_{e}$/ $\sigma_{c}$ $>$ $H_{c}$.
\end{itemize}

Within at most 1 kpc both $H_{c}$ and $H_{e}$ are assumed constant and regulated
by the disk force per unit mass, $K_{Z}$.

In this Section, the gravitational effects of a central BH (and the related Bulge mass)
are described.
Down to where, along $R$, it can be assumed (lacking obvious reasons to the contrary) 
that $\sigma_{c}$ and $T_{e}$ remain constant (as well as the Bulge density), 
this addition shall result into a change in $H_{c}$,
namely this quantity will decrease with decreasing $R$, 
as soon as the vertical attraction exerted by the
BH will start dominating. The profile $H_c(R)$ will go through an inflection
at $R_{infl}$, which regulates the cone opening angle, and increases with $M_{BH}$. Thus,
if there were a statistical correlation between
$M_{BH}$ and $L_{x}$, the consequence would be a trend of the type given in Sect.~2,
which is encouraging.

The force per unit mass exerted in the vertical, $Z$, direction by the matter in the disk
can be expressed, within about Z=100 pc, as 

 \begin{equation}
K_{Z,d}=-4\pi G \rho_{d} Z
\end{equation}
\noindent
where $\rho_{d}$ is the approximately constant density of the matter in the disk, mainly 
provided by stars.

The equivalent force exerted by the BH, in the approximation $Z\ll R$, is given by

\begin{equation}
K_{Z,BH}=-G\frac{M_{BH}}{R^{3}}Z
\end{equation}

Similarly, the force exerted by the Bulge is 

\begin{equation}
K_{Z,b}=-G\frac{M_b(R)}{R^{3}}Z = -G\frac{4\pi\rho_b}{3} Z
\end{equation}

\noindent 
where $M_b(R)$ is the bulge mass within the radius $R$.  We have assumed 
a spherical symmetry, and a costant bulge density $\rho_b$ within $R$ 
(see the Appendix, where $\rho_b$ is calculated as a function of the Black Hole mass).

From the equation that links the clouds velocity dispersion and scale height to the
sum of the three contributions to $K_Z$

\begin{equation}
\sigma^{2}=2\int_{H_{c}}^{0}(K_{Z,d}+K_{Z,BH}+K_{Z,b})dZ
\end{equation}

\noindent
one infers

\begin{equation}
H_{c}(R)=\left(\frac{\sigma^{2}}{4\pi G \rho_{d}+\frac{GM_{BH}}{R^{3}}+ 
\frac{4\pi G \rho_b}{3} }\right)^{1/2} .
\label{Hc}
\end{equation}

The distance of the point of inflection is then immediately derived to be

\begin{equation}
R_{infl}=\frac{1}{2}\left(\frac{M_{BH}}{4\pi \rho_{d}+ \frac{4\pi\rho_b}{3}}\right)^{1/3} .
\end{equation}

\begin{figure}[h]
\begin{center}
\includegraphics[width=8 cm]{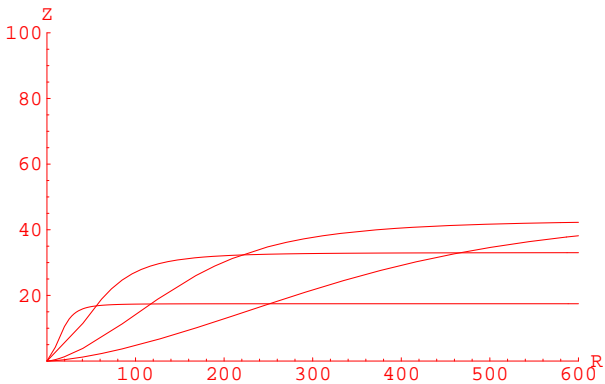}
\caption{Scale height of the clouds as a function of R when $M_{BH}$=$10^{6}$ $M_{\odot}$,
$10^{7}$ $M_{\odot}$, $10^{8}$ $M_{\odot}$ and $10^{9}$ $M_{\odot}$.}
\label{flesso}
\end{center} 
\end{figure}

To be noted that at $R$=2$R_{infl}$ the gravitational pull of the BH in the $Z$ direction 
is equal to that of the matter in the disk and the bulge.
For values of $R$ much lower than 2$R_{infl}$, where the BH dominates, we shall
disregard the contribution of the shrinking disk gravity 
(that is the increase of $\rho_{d}$ with
decreasing $R$),
and the profiles of $H_c$ shown in Figure~\ref{flesso} correspond to a choice of 
$\rho_{d}$ equal
to the value typical of a galaxy like our own, namely  0.85 solar masses per cubic parsec,
a velocity dispersion $\sigma$ = 10 km $s^{-1}$, and four values of $M_{BH}$ (and thence
of $\rho_b$). The opening angle is very sensitive to the last
parameter, and depends on the value of $R_{infl}$
increasing from 12 pc when $M_{BH}$ = 10$^6$ $M_{\odot}$ to 226 pc when 
$M_{BH}$ = 10$^9$$M_{\odot}$.

In order to evaluate the effective obscuration, this geometrical effect must be accompanied
by assumptions on the number of clouds along a line of sight. To this end, the
vertical distribution of the clouds number density is assumed to follow an exponential law

\begin{equation}
n(Z, R)=n_{0}(R)e^{-({Z \over H_{c}(R)})}
\label{}
\end{equation}

\noindent
where $H_c(R)$ is given by eq. \ref{Hc} 
and $n_0$ is defined through the surface density $\Sigma$,
assumed constant with $R$:

\begin{equation}
\Sigma = 2 \!M_{c} \!\!\! \int_{0}^{\infty} \! n_{0}(R)e^{-({Z \over H_{c}(R)})} dZ
\end{equation}

\noindent
where M$_c$ is the mass chosen to represent the single cloud.
Thus:

\begin{equation}
n(Z, R)=n_{0}\frac{H_{c}}{H_{c}(R)}e^{-({Z \over H_{c}(R)})},
\label{densnubi}
\end{equation}

\noindent
where $H_c$ is the scale height at $R\gg R_{infl}$.
Except for the parameter $\Sigma$, all the other adopted quantities 
are given in Table 1. In this table also a tidal radius is given, 
namely the distance within which
the single cloud would be disrupted by the tidal force exerted by the BH, namely

 \begin{equation}
R_{tidal}=\left(\frac{3M_{BH}}{2\pi \rho_{c} m_H}\right)^{1/3},
\end{equation}

\noindent
where $\rho_{c}$ is the internal density of the cloud, 
assumed equal to 10${^4}$ hydrogen molecules 
cm$^{-3}$. It should be noted that 
$R_{tidal}$ is of the same order of $R_{infl}$ for a BH mass of
10$^6$ $M_{\odot}$, and lower for higher masses. 
The adopted cloud size is such that the corresponding column density is equal 
to 10$^{22}$ cm$^{-2}$. Because this is the value of N$_H$ chosen to discriminate
between Seyferts with no absorption and Compton-thin absorbed ones, we have then
simply to calculate
the probability that one single cloud happens to lie along the line of sight. 
It is easy to
demonstrate that it does not make a difference if such clouds, or smaller ones, where congregated
in complexes rather than dispersed, as assumed, provided that the 
number of complexes remains large.

\begin{table} [h]
\begin{center}
{
\begin{tabular}{|l|l|}
\hline  disk matter density & $\rho_{d}= 0.84 M_{\odot} pc^{-3}$    \\
\hline clouds velocity dispersion & $\sigma$=10 km/s \\
\hline scale height of the clouds & $H_{c}$=50 pc\\
\hline typical cloud inner density & $\rho_{c}= 10^{4}cm^{-3}$ \\
\hline typical cloud radius  & $r_{c}$=0.167 pc \\
\hline typical cloud column density & $N_{H,c}=10^{22}cm^{-2}$ \\
\hline typical cloud mass & $M_{c}=4.35 M_{\odot}$\\
\hline typical cloud temperature & $T_{c}=10$ K\\
\hline external gas temperature  & $T_{e}=10^6$ K \\
\hline external gas density  & $\rho_{e}=0.1$ cm$^{-3}$ \\
\hline cloud number density at z=0 & $n_{0}=\Sigma/ 2 H_{c}M_{c}$ \\
\hline tidal radius R for  $M_{BH}=10^{9} M_{\odot}$   & $R_{tidal}$=128 pc \\
\hline tidal radius R for  $M_{BH}=10^{8} M_{\odot}$   & $R_{tidal}$=60 pc \\
\hline tidal radius R for $M_{BH}=10^{7} M_{\odot}$  & $R_{tidal}$=28 pc\\
\hline tidal radius R for $M_{BH}=10^{6} M_{\odot}$ & $R_{tidal}$=13 pc \\
\hline inflection radius R for  $M_{BH}=10^{9} M_{\odot}$   & $R_{infl}$=226 pc \\
\hline inflection radius R for  $M_{BH}=10^{8} M_{\odot}$   & $R_{infl}$=101 pc \\
\hline inflection radius R for $M_{BH}=10^{7} M_{\odot}$  & $R_{infl}$=39 pc\\
\hline inflection radius R for $M_{BH}=10^{6} M_{\odot}$ & $R_{infl}$=12 pc \\
\hline 
\end {tabular}
}
\\
\caption {Adopted values for the parameters. See text for details.}
\end{center}
\end {table}

The line of sight is defined through the angle $\theta$ given by

\begin{equation}
\frac{Z}{R}=tg\theta,
\end{equation} 

\noindent
and the number density as a function of $\theta$ and R obtains from eq. \ref{densnubi}:

\begin{equation}
n(\theta, R)=n_{0}\frac{H_{c}}{H_{c}(R)}e^{-({R tg\theta \over {H_{c}(R)} })}
\end{equation}

For a given $\theta$ the number of encountered clouds is given by

\begin{equation}
N=\pi r_{c}^{2}\int_{R_{in}}^{R_{out}} n(\theta, R) dR,
\end{equation}

\noindent
where $\pi$ $r_{c}^{2}$ is the geometric cross section of a cloud, $R_{out}$=4$H_{c}$/tg$\theta$ 
and $R_{in}$ is consistently adopted equal to $R_{tidal}$.

\begin{figure}[h]
\begin{center}
\includegraphics[width=8 cm]{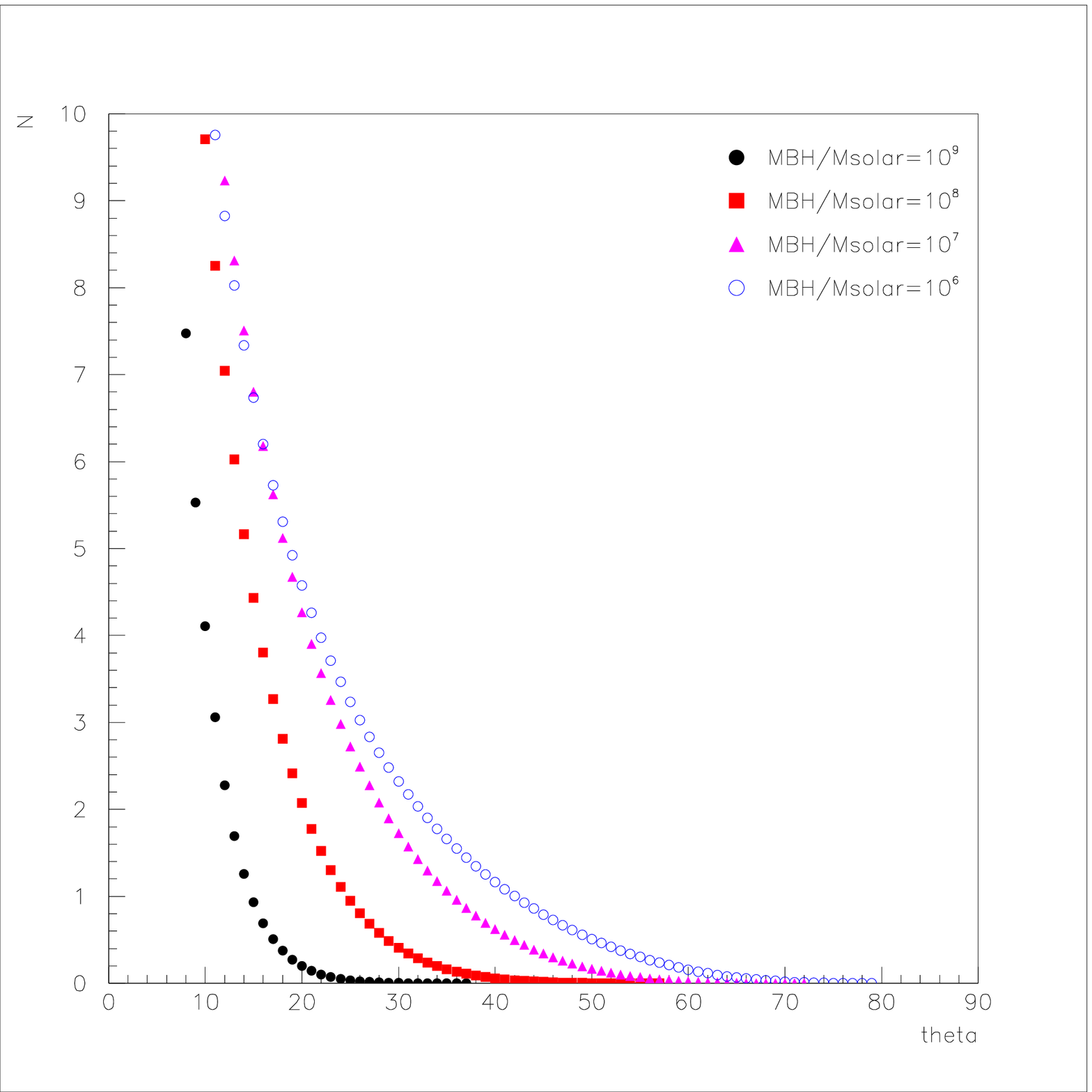}
\caption{Number of clouds along the line of sight as a function of $\theta$ for four
 values of M$_{BH}$,
when $\Sigma$=200 $M_{\odot}$ pc$^{-2}$}\label{spessoreottico}
\end{center} 
\end{figure}

Figure \ref{spessoreottico} illustrates the number of clouds encountered 
as a function of $\theta$, for four values of the $M_{BH}$ and $\Sigma$ equal to 200 $M_{\odot}$ pc$^{-2}$. 

By defining $\theta_{\star}$ as the angle where this number equals 1, it is straightforward
to estimate the solid angle, hence the fraction of sources $N_{s}$ where the line of sight is
bound to cross a column of at least 10$^{22}$ cm$^{-2}$,

\begin{equation}
\frac{N_{s}(log N_{H}>22)}{N_{s,tot}}=cos(90-\theta_{\star})
\end{equation} 

\noindent
as a function of two parameters,
the BH mass and the surface density $\Sigma$ of the molecular gas within 
$R_{out}$.

\begin{figure*}
\hbox{
\includegraphics[width=8 cm]{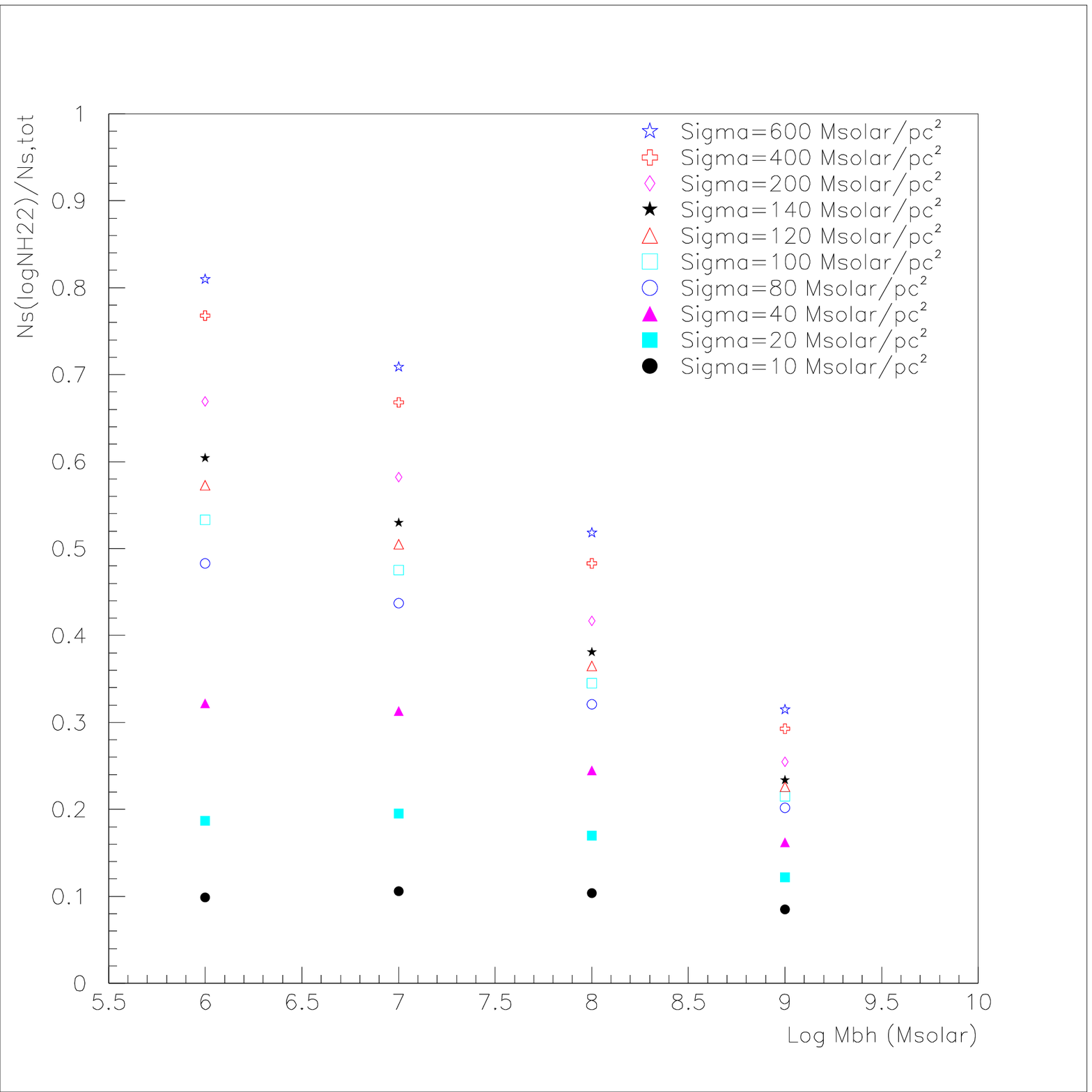}
\hspace{1.5cm}
\includegraphics[width=8 cm]{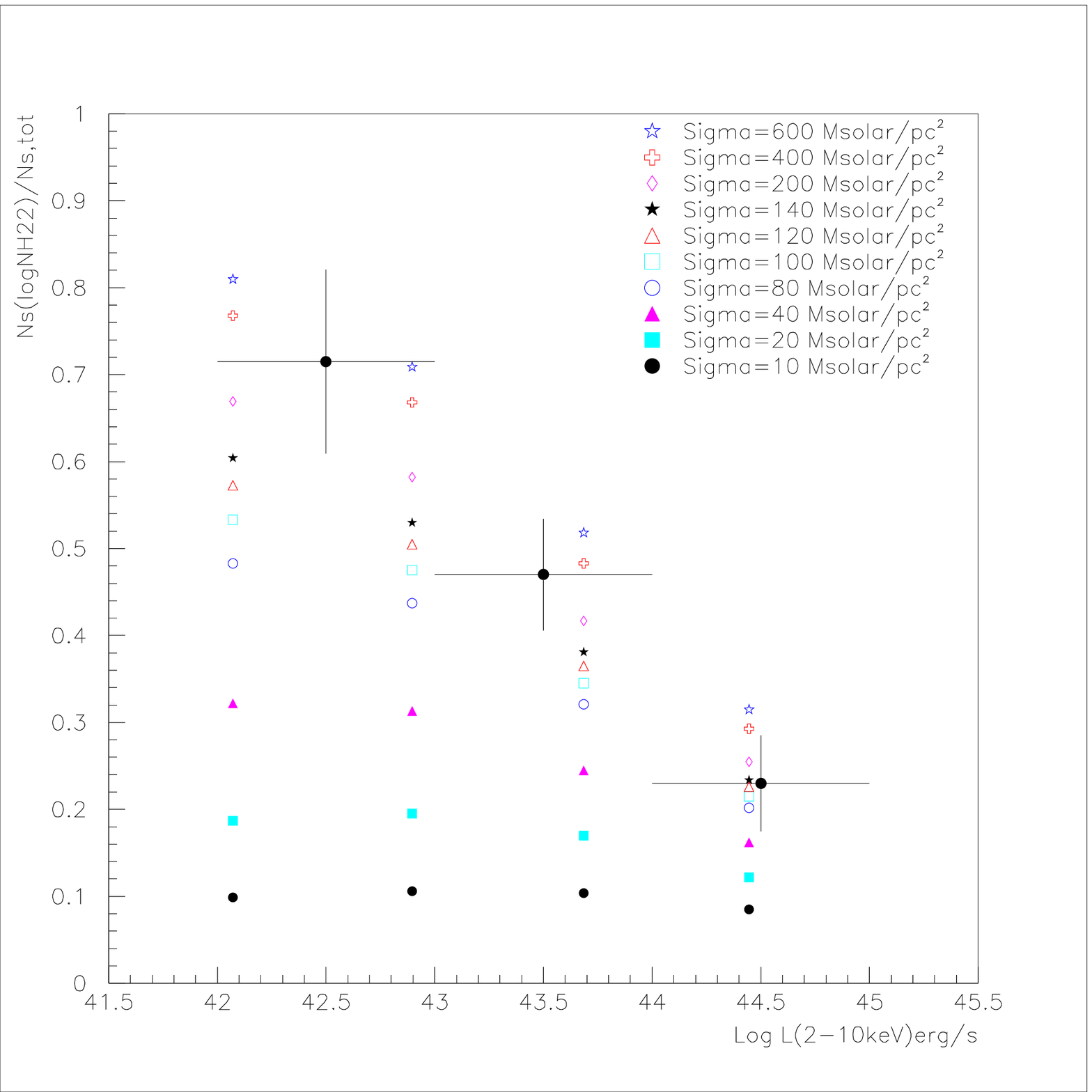}
}
\caption{Expected behaviour of the fraction N$_{s}$(logN$_{H}$ $>$ 22)/N$_{s,tot}$, 
for several values of $\Sigma$, as a function $M_{BH}$ (left hand panel) and
of L(2-10 keV) (right hand panel), with the latter derived assuming  $L_{Bol}$/$L_{Edd}$=0.1
(e.g. Peterson et al. 2004) 
and the luminosity--dependent bolometric correction in Marconi et al. (2004). In the right hand 
panel, crosses represent the observed fraction (corrected for selection effects, and adapted from 
La Franca et al. 2005) of obscured AGN in the local universe.}
\label{anticorr}
\end{figure*}

In Figure~\ref{anticorr} (left panel), the fraction of absorbed AGN is shown for four
different values of the BH mass (10$^6$, 10$^7$, 10$^8$  and 10$^9$ $M_{\odot}$) and
several choices of $\Sigma$. 
In order to draw a comparison with the observational evidence, where the observable is
the X-ray luminosity, it is necessary to adopt a relation between $L_x$ and 
$M_{BH}$.  The luminosity-dependent bolometric correction in Marconi et al.
(2004; see their eq.(21) and Fig.~3) and a luminosity at 0.1 times the  
Eddington limit (see e.g. Peterson et al. 2004) have been adopted.
The results are given in Figure~\ref{anticorr} (right panel), 
where the best-fit relationship between the fraction of absorbed AGN and X-ray luminosity
obtained by La Franca et al. (2005) is shown for comparison.

The dependence of the slope of
the anticorrelation on the adopted value of $\Sigma$ is noteworthy. 
While for values of this parameter
larger than about 80-100 $M_{\odot}$ pc$^{-2}$ the slope changes rather little, and remains
gratifyingly close to the observed one, for smaller values it becomes shallower, and
the more so the lower the value. This comes out simply because, in order to intercept a cloud,
one needs to integrate well beyond the inflection radius: in other words, despite the shrinking
in $H_c$ due to the black hole, the number density remains too small to have any sizeable effect
within 2$R_{infl}$, and the required absorption is obtained only for line of sights grazing the
plane of the disk.

\section{Discussion and conclusions}

The main result of this paper is that the anticorrelation between  $\xi$(L$_x$)
(namely the fraction of AGN with N$_H$, as measured by 
their X-ray spectra, greater than 10$^{22}$ cm$^{-2}$) and 
L$_x$ in Compton-thin objects is at least qualitatively reproduced
provided that: a) there is a sufficiently large number of molecular clouds
within the radius where the BH gravitational influence is dominant; 
b) there is, in a statistical sense, a correlation between the BH mass
and the AGN luminosity. These two conditions are now briefly discussed.

As shown in Fig.~\ref{anticorr}, the slope and the normalization of the 
anticorrelation are consistent with the observational
results for values of $\Sigma \geq$ 150-200 $M_{\odot}$ pc$^{-2}$, if a bolometric luminosity 0.1 times the Eddington 
luminosity is assumed (as it seems to be the case in the local universe,
e.g. Peterson et al. 2004). Thus, the first question arising is what are the
typical values of this quantity in the innermost region of spiral galaxies 
(the hosts of Seyfert active nuclei). The relevant radius of this region,  
2$R_{infl}$, lies between about 25 and 450 pc (for $M_{BH}$ from one and 
one thousand million solar masses), which corresponds to 0.5-9 arcsec at a distance
of 10 Mpc, and requires the best angular resolution currently achievable,
with interferometry at the 3 mm CO J=1-0 line, on relatively nearby galaxies.
In this respect, the BIMA Survey of Nearby Galaxies (BIMA SONG) at Hat Creek, CA,
with the 10-element Berkeley-Illinois-Maryland Association (BIMA)
millimeter interferometer (Helfer et al. 2003) is the best available for
our purpose, because the sample was deliberatly selected without reference
to CO or infrared brightness. It consists of 44 spiral galaxies of all
morphological types (the Sa type are somewhat undersampled relative
to the later types). The angular resolution of 6 arcsec corresponds to
360 pc (that is 180 pc radius for the central region) at the average
distance of the galaxies, 12 Mpc. From the CO surface brightness distribution
(see Table 5 in Helfer et al. 2003) it turns out that the peak brightness
is attained within the central 6 arcsec in 20 out of 44 galaxies. In these
20 objects the corresponding surface density $\Sigma_{centr}$ lies between
15 and 1854 $M_{\odot}$ pc$^{-2}$, with most of them (18) above 90, and the
other two around 15 $M_{\odot}$ pc$^{-2}$. Notably, among the other 24
galaxies, 10 have $\Sigma_{centr}$ $\geq$ 50 $M_{\odot}$ pc$^{-2}$, with values
up to 1151. The galaxies with $\Sigma_{centr}$ $\geq$ 150 
$M_{\odot}$ pc$^{-2}$ are 22/44, namely 50\%. The straight mean value of  this 22
galaxies is 511 $M_{\odot}$ pc$^{-2}$, 
notably close to the value of $\sim$500 $M_{\odot}$ pc$^{-2}$ of the 300 pc molecular
disk in our own galaxy (G\"usten 1989). 

The sample contains 8 objects classified as 
Seyfert galaxies, 6 of them with $\Sigma_{centr}$ greater than 90 and up to 466
$M_{\odot}$ pc$^{-2}$, 1 (NGC 4725) with $\Sigma_{centr}$ = 21 $M_{\odot}$ pc$^{-2}$,
1 (NGC3031) with $\Sigma_{centr}$ below the detection limit: the proportion of
galaxies with an active nucleus, which possess a high value of $\Sigma_{centr}$,
appears to be, within the small statistics, similar to that in the general population.

The fraction of objects which meet our requirements is encouraging, 
especially because it does not require a special correlation with the nuclear
activity, and represents a relatively large scale (compared to the close circumnuclear
environment) property fairly common in spiral galaxies. In other words,
the condition required is already in place in about half of the galaxies
before a high rate nuclear activity lights up, which in turn is unlikely
to significantly influence the preexisting condition. Much higher resolution and
sensitivity surveys are needed to explore more distant objects, and therefore
draw more solid conclusions on the validity of our hypothesis: to pursue this
goal an array of millimeter telescopes like ALMA is necessary.

Concerning the increase of the fraction of absorbed sources with 
redshift (La Franca et al. 2005), 
an interesting possibility is that it might be
due to an average increase of $\dot{M}$ with the redshift,
if one were allowed to assume that the molecular content and distribution
in spiral galaxies is independent of $z$. 
Indeed, McLure \& Dunlop (2004) found that the 
$L_{bol}/L_{Edd}$ ratio raises from about 0.15 at $z\sim$0.2 to 0.5 at  $z\sim$2.
We intend to investigate this issue further,
in particular within the frame of astrophysically self consistent
models, which associate the growth of the supermassive BH in galaxies
with the evolution in cosmic time of the AGN population (e.g. Menci et al. 2004).

\section*{Acknowledgements}
The authors are grateful to Renzo Sancisi and Roberto
Maiolino for useful discussions and for their help in exploring the literature on the CO central 
concentration in galaxies. They thank the referee for his constructive comments. 
They acknowledge financial support from ASI and MIUR (under
grant {\sc prin-02-02-23}).

\appendix
\section{Effects of the stellar bulge}

The BH mass is known to correlate with the velocity dispersion, $\sigma_b$, in the bulge:

\begin{equation}
\frac{M_{BH}}{10^{8}M_{\odot}}=(1.66\pm0.24)\left(\frac{\sigma_b}{200 Km s^{-1}}\right)^{\alpha},
\label{Msigma}
\end{equation} 

\noindent
with $\alpha = 4.86\pm0.43$ (Ferrarese \& Ford 2005).

With respect to the virial mass of the bulge 
\begin{equation}
 GM_{b}\approx \sigma^{2} R_{b},
\label{virialbulgemass}
\end{equation}

\noindent
the BH mass turns out (Merritt $\&$ Ferrarese 2001) proportional to M$_b$, M$_{BH} \simeq 10^{-3}$M$_b$.
From eq. \ref{Msigma} and \ref {virialbulgemass} one immediately obtains

\begin{equation}
R_{b} \propto \frac{M_{b}}{\sigma_{b}^{2}} \propto 
\frac{M_{BH}}{M_{BH}^{2/\alpha}}\propto M_{BH}^{1-2/\alpha},
\end{equation}
or, rounding off $\alpha$ to 5,
\begin{equation}
R_{bulge} \propto M_{BH}^{3/5}.
\end{equation}

Adopting for the density profile of the bulge the relationship:

\begin{equation}
\rho(R)=\frac{\rho_{b}}{1+(R/R_{c})^{2}}
\end{equation}

\noindent
where $\rho_{b}$ is the central density and $R_{c}$ is the core radius,
if it is assumed that  $R_{c}$ is proportional to  $R_{b}$, it follows
that $R_{c} \propto M_{BH}^{3/5}$. After normalization
to the values which apply to the Milky Way, namely $R_{c,MW}$=400 pc, 
$M_{BH,MW}$=$4\times 10^{6} M_{\odot}$, one obtains:

$
\begin{array}{cc}  
M_{BH}=10^{6}M_{\odot} & R_{c}=180~pc   \\ 
M_{BH}=10^{7}M_{\odot} & R_{c}=700~pc \\ 
M_{BH}=10^{8}M_{\odot} & R_{c}=2.7~Kpc \\
M_{BH}=10^{9}M_{\odot} & R_{c}=11~Kpc
\end{array}
$

\noindent
For these values of M$_{BH}$ $R_c$ turns out to be much larger than two times
$R_{infl}$ defined in Sect.~4  and given in Table 1. Thus it can be concluded
that the stellar bulge density within 2$R_{infl}$ remains practically constant.
Its value, in the above approximations, is proportional to $M_{BH}^{-4/5}$,
and we calculated it for each BH mass by simply rescaling the Milky Way value:

$
\begin{array}{cc}  
M_{BH}=10^{6}M_{\odot} & \rho_{b}=15.1~M_{\odot}/pc^3   \\
M_{BH}=10^{7}M_{\odot} & \rho_{b}=2.4~M_{\odot}/pc^3 \\ 
M_{BH}=10^{8}M_{\odot} & \rho_{b}=0.4~ M_{\odot}/pc^3 \\
M_{BH}=10^{9}M_{\odot} & \rho_{b}=0.06~M_{\odot}/pc^3
\end{array}
$
\end{document}